\documentclass{gretsi}

\usepackage[english,french]{babel}    
\usepackage{times}			
\usepackage[utf8]{inputenc}
\usepackage[numbers, square, comma, compress]{natbib}
\usepackage{setspace}
\usepackage{fancyhdr}
\usepackage{dsfont}
\usepackage{amsfonts}
\usepackage{amssymb}
\usepackage{amsthm}
\usepackage{amsmath}
\usepackage{mathrsfs}
\usepackage{xcolor}
\usepackage{graphicx}
\usepackage{mathdots}
\usepackage{float}
\usepackage[colorinlistoftodos]{todonotes}
\usepackage[ruled,vlined]{algorithm2e}
\usepackage{verbatim}
\usepackage{tikz}
\usepackage[font={small}]{caption}
\usetikzlibrary{calc}

\theoremstyle{theorem}
\newtheorem{teo}{Theorem}

\newtheorem{lem}[teo]{Lemma}

\newtheorem{oss}[teo]{Remark}
\newtheorem{defi}[teo]{Definition}

\theoremstyle{definition}

\DeclareMathOperator{\V}{\mathcal V}
\DeclareMathOperator{\C}{\mathcal C}
\DeclareMathOperator{\X}{\mathcal X}

\title{Polar codes for empirical coordination over noisy channels with strictly causal encoding}

\author{\coord{Giulia}{Cervia}{1},
        \coord{Laura}{Luzzi}{1},
    \coord{Ma\"{e}l}{Le Treust}{1},
    \coord{Matthieu R.}{Bloch}{2}}

\address{\affil{1}{ETIS UMR 8051, Université Paris Seine, Université Cergy-Pontoise, ENSEA, CNRS, Cergy, France.}
         \affil{2}{School of Electrical and Computer Engineering, Georgia Institute of Technology, Atlanta, Georgia}}

\email{\{giulia.cervia, laura.luzzi, mael.le-treust\}@ensea.fr, matthieu.bloch@ece.gatech.edu}

\frenchabstract{
Dans ce travail, nous proposons un schéma de codage basé sur les codes polaires pour la coordination 
empirique d’appareils autonomes. Nous considérons un réseau simple composé de deux nœuds reliés par un 
lien bruité, et nous cherchons à coordonner les signaux en entrée et en sortie du canal, avec la source et sa reconstruction. Lorsque l'encodeur est strictement causal, nous montrons que les codes polaires atteignent la région optimale de coordination empirique, à condition que les deux nœuds partagent une source aléatoire, dont le débit est asymptotiquement négligeable. }

\englishabstract{In this paper, we propose a coding scheme based on polar codes for empirical coordination of autonomous devices.
We consider a two-node network with a 
noisy link in which the input and output signals have to be coordinated with the source and 
the reconstruction. 
In the case of strictly causal encoding, we show that polar codes achieve the empirical coordination region, provided that a vanishing rate of common randomness is available.}

\begin{document}\sloppy
\maketitle

\section{Introduction}
\vspace{-0.35cm}
In decentralized networks of connected objects,
such as wireless sensors, medical and wearable devices, smart energy meters, 
home appliances, and self-driving cars, devices sense their environment and choose their actions in order 
to achieve a general objective.
It is essential that these devices, considered as autonomous decision-makers, cooperate and coordinate their actions to induce a global behavior, represented by a utility function to be maximized.

Within the framework of information theory, two different metrics have been proposed to measure the level of 
coordination:  \emph{empirical coordination} requires the joint histogram of the actions to approach a target distribution, 
while \emph{strong coordination} requires the joint distribution of actions to converge in total variation to an i.i.d. target distribution \cite{cuff2010}. 

We consider a two-node network with an information source and a noisy channel in which the input and output signals should be empirically coordinated with the source and the reconstruction. 
In \cite{cuff2011hybrid}, the authors provide a characterization of the \emph{coordination region} when the encoder is strictly causal. 
Inspired by the binning technique using polar codes in \cite{chou2015polar}, we propose an explicit coding scheme that achieves a subset of the coordination region in \cite{cuff2011hybrid} by turning the argument of \cite{choundhuri2010capcity} into an explicit polar coding proof. 
The scenario in which both the encoder and the decoder are non-causal has already been considered for empirical  coordination
with polar codes \cite{Cervia2016}. Here, we focus on the setting in which the encoder is strictly causal.

In this paper, we only achieve a subset of the coordination region because of the use of binary polar codes, but the whole region can be achieved using non-binary polar codes. 

The remainder of the paper is organized as follows. $\mbox{Section \ref{sec:problem}}$ introduces the notation, describes the model under investigation and states the main achievability result. $\mbox{Section \ref{sec:polar}}$ details the proposed coordination scheme using polar codes. 
Finally, Section \ref{subsec:coord} proves the main result.
\vspace{-0.5cm}
\section{Problem statement}\label{sec:problem}
\vspace{-0.3cm}
\subsection{Notation}\label{sec: not}
We define $[a,b]$ as the set of the integers between $a$ and $b$.
For $n=2^m$, $m \in \mathbb N$, we note $G_n:= \begin{footnotesize}
\begin{bmatrix}
1 & 0\\
1 & 1
\end{bmatrix}^{\otimes m}
\end{footnotesize}$ the source polarization transform defined in \cite{arikan2010source}. 
Given $X^{1:n}\!:= \!(X^1\!, \ldots, X^n)$ a random vector, we note $X^{1:j}$ the first $j$ components of $X^{1:n}$ and $X[A]$, where $A\! \subset\! [1,n]$, the components $X^j$ such that $j\! \in \!A$.
We note $\mathbb V (\cdot , \cdot)$ and $\mathbb D (\cdot \Arrowvert \cdot)$ the variational distance and the Kullback-Leibler divergence between two distributions, respectively. 
\vspace{-0.4cm}
\subsection{System model and main result}\label{sec: model}
\vspace{-0.2cm}
\begin{center}
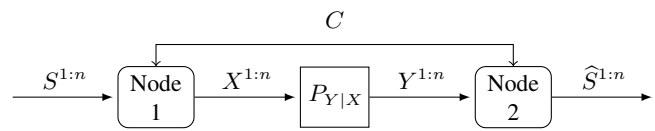
\begin{figure}[h]
\begin{small}
\begin{tikzpicture}[
nodetype1/.style={
	rectangle,
	minimum width=7mm,
	text width=7mm,
	text centered,
	minimum height=8mm,
	draw=black
},
nodetype2/.style={
	rectangle,
	rounded corners,
	minimum width=8mm,
	text width=8mm,
	text centered,
	minimum height=8mm,
	draw=black
},
tip2/.style={-latex,shorten >=0.6mm}
]
\matrix[row sep=0.4cm, column sep=1.4cm, ampersand replacement=\&]{
\node ( ) {}; \& 
\node ( ) {}; \& 
\node (C) {$C$}; \& \\
\node (PS)  {}; \& 
\node (uno) [nodetype2, text centered] {Node \\ 1}; \& 
\node (PYX) [nodetype1] {$P_{Y|X}$}; \& 
\node (due) [nodetype2, text centered] {Node \\ 2}; \&
\node (PST) {}; \\};
\draw[<->] (uno)  -- ++ (0,0.7) -|  (due) ;
\draw[->] (PS) edge[tip2] node [above] {$S^{1:n}$} (uno) ;
\draw[->] (uno) edge[tip2] node [above] {$X^{1:n}$} (PYX) ;
\draw[->] (PYX) edge[tip2] node [above] {$Y^{1:n}$} (due) ;
\draw[->] (due) edge[tip2] node [above] {$\widehat S^{1:n}$} (PST) ;

\end{tikzpicture}
\end{small}
\vspace{-0.4cm}
\caption{Coordination of signals and actions for a two-node network with a noisy channel.}
\label{fig: coord}
\end{figure}
\end{center}
\vspace{-0.8cm}
We consider two agents, Node 1 and Node 2, who have access to a shared randomness source $C \in \mathcal C_n$ (Figure \ref{fig: coord}). 
Node 1 observes an i.i.d. sequence of actions  $S^{1:n} \in \mathcal S^n$ with discrete probability distribution $P_S$. 
Node 1 then selects a signal $X^{1:n}$ such that $X^{i}= f_i(S^{1:i-1}, C)$, where $f^n=\{f_i\}_{i=1}^n$, $f_i: \mathcal S^{i-1} \times  \mathcal C_n \rightarrow \X$ is the strictly causal encoder. 
The signal $X^{1:n}$ is transmitted over a discrete memoryless channel with transition probability $P_{Y |X}$.
Upon receiving $Y^{1:n} \in \mathcal Y^n $, Node 2 selects an action $\widehat S^{1:n} = g^n(Y^{1:n}, C)$, where $g^n: \mathcal Y^n  \times  \mathcal C_n \rightarrow \widehat{\mathcal S}^n$ is the non-causal decoder.
For block length $n$, the pair $(f^n , g^n )$  constitutes a code.
Node 1 and Node 2 wish to coordinate in order to obtain a joint distribution of actions and signals that is close to a target distribution $P_{SXY\widehat{S}}$.
We focus on the empirical coordination metric defined in \cite{cuff2010}.
\begin{defi}
A distribution $P_{SXY\widehat{S}}$ is achievable if for all 
$\varepsilon >0$ there exists a sequence of codes $\{(f^n , g^n )\}_{n \in \mathbb N}$ 
such that
$$\lim_{n\rightarrow \infty}\mathbb P \left\{ \mathbb V \left(  T_{S^{1:n} X^{1:n} Y^{1:n} \widehat{S}^{1:n}},P_{SXY\widehat{S}} \right) > \varepsilon \right\} =0,$$
where $T_{S^{1:n} X^{1:n} Y^{1:n} \widehat{S}^{1:n}}(s,x,y,\hat s)$ 
is the empirical distribution of the tuple $(S^{1:n}, X^{1:n}, Y^{1:n}, \widehat{S}^{1:n})$ 
induced by the code.

The empirical coordination region $\mathcal C$ is the set of achievable distributions $P_{SXY\widehat{S}}$.
\end{defi}

\begin{teo}[Strictly causal encoder \cite{cuff2011hybrid}] \label{theo3}
Let $P_{S}$ and $P_{Y|X}$ be the given source and channel parameters. When the encoder is strictly causal,
 the coordination region $\C$ is given by
 \begin{equation}\label{eq: c}
 \C:= \begin{Bmatrix}
P_{SXY\widehat{S}} :\mbox{ } 
P_{SXY\widehat{S}}=  P_{S}  P_{X}  P_{Y|X}  P_{\widehat{S}|SXY}  \\
\mbox{ }\exists \mbox{ } U \mbox{ taking values in $\mathcal U$}\\ 
 P_{SXYU \widehat S}= P_S P_{X} P_{U|XS} P_{Y|X} P_{\widehat{S}|UY}\\
\mbox{ } I(X,U;S) \leq I(X,U;Y)\\
\mbox{ } \lvert \mathcal U \rvert \leq \lvert \mathcal S \rvert \lvert \mathcal X \rvert \lvert \mathcal Y \rvert  \lvert \widehat{\mathcal S} \rvert+1 \\
\end{Bmatrix}
\end{equation}
 \end{teo}

 \begin{oss}
By the chain rule, we have
\begin{itemize}
\item[\textbullet]$I(X,U;S)= I(U;S|X)+I(X;S)=I(U;S|X)$ since $S^{1:n}$ and $X^{1:n}$ are independent;
\item[\textbullet] $I(X,U;Y)= I(U;Y|X)+I(X;Y)=I(X;Y)$ because of the Markov chain $U-X-Y$.
\end{itemize}
Hence the condition $I(X,U;S) \leq I(X,U;Y)$ in \eqref{eq: c} becomes $I(U;S|X)\leq I(X;Y)$.
\end{oss}

\begin{teo}\label{theo}
For all $P_{SXY\widehat{S}} \in \mathcal C$ such that  $\mathcal U = \{ 0,1\}$,
there exists an explicit polar coding scheme that achieves empirical coordination with vanishing rate of common randomness.
\end{teo}

\begin{oss}
Since $U$ is binary we only achieve a subset of $\mathcal{C}$. The proof can be generalized to the case where $\lvert \mathcal U \rvert$ is a prime number using non-binary polar codes. 
\end{oss}

\vspace{-0.8cm}
\section{Polar coding scheme}\label{sec:polar}
\vspace{-0.2cm}
Consider the random vectors $S^{1:n}$, $U^{1:n}$, $X^{1:n}$, $Y^{1:n}$ and $\widehat S^{1:n}$
generated i.i.d. according to $P_{SXUY \widehat S}$ that factorize as in \eqref{eq: c} with the same mutual information and cardinality constraints.

\vspace{-0.3cm}
\paragraph{Polarize $X$}
Let $Z^{1:n}=X^{1:n}G_n$ be the polarization of $X^{1:n}$, where $G_n$ is the source polarization transform.
For some $\mbox{$0<\beta<1/2$}$, let $\delta_n: = 2^{-n^ {\beta}}$ and define the very high and high entropy sets:
{\allowdisplaybreaks
 \begin{align*}
\mathcal V_{X}: & =\left\{j\in[1,n]:H(Z^j|Z^{1:j-1})>1-\delta_n \right\},\\
\mathcal H_{X}: & =\left\{j\in[1,n]:H(Z^j|Z^{1:j-1})>\delta_n \right\},  \stepcounter{equation}\tag{\theequation}\label{eq: hz}\\
\mathcal H_{X | Y}: & =\left\{j\in[1,n]:H(Z^j|Z^{1:j-1} Y^{1:n})>\delta_n \right\} .
 \end{align*}}
Partition the set $[1,n]$ into four disjoint sets:
\begin{align*}
A_1 := \mathcal V_{X} \cap \mathcal H_{X|Y} , \quad & A_2 : = \mathcal V_{X} \cap \mathcal H_{X|Y}^c,\\
A_3 := \mathcal V_{X}^c \cap \mathcal H_{X|Y} ,\quad
& A_4 := \mathcal V_{X}^c \cap \mathcal H_{X|Y}^c.
\end{align*}

\begin{oss}\label{oss card1}
We have:
\begin{itemize}
\item[\textbullet] $\mathcal V_{X} \subset \mathcal H_{X}$ and $\displaystyle \lim_{n \rightarrow \infty} \frac{\lvert \mathcal H_{X} \setminus \mathcal V_{X}  \rvert}{n} = 0$ \cite{arikan2010source},
 \item[\textbullet] $A_1 \cup A_2 = \mathcal V_X$ and $\displaystyle \lim_{n \rightarrow \infty} \frac{\lvert \mathcal V_{X} \rvert}{n}  = H(X)$  \cite {chou2015secretkey},
 \item[\textbullet]  $A_1 \cup A_3 = \mathcal H_{X|Y}$ and $\displaystyle \lim_{n \rightarrow \infty} \frac{\lvert \mathcal H_{X | Y} \rvert}{n} = H(X|Y)$ \cite{arikan2010source}.
\end{itemize}
Since  $\displaystyle \lim_{n \rightarrow \infty} \frac{\lvert A_2\rvert - \lvert A_3 \rvert}{n}= H(X)- H(X|Y) = I(X;Y)\geq 0$ 
this implies directly that for $n$ large enough $\lvert A_2 \rvert \geq \lvert A_3 \rvert$.
\end{oss}

\vspace{-0.5cm}
\paragraph{Polarize $U$}
Let $V^{1:n}=U^{1:n}G_n$ be the polarization of $U^{1:n}$ and define:
{\allowdisplaybreaks
 \begin{align*}
\mathcal V_{U|XS}: & =\left\{j\in[1,n]:H(V^j|V^{1:j-1}X^{1:n}S^{1:n})>1-\delta_n \right\},\\
\mathcal H_{U | XS}: & =\left\{j\in[1,n]:H(V^j|V^{1:j-1} X^{1:n}S^{1:n})>\delta_n \right\}, \\
\mathcal H_{U | X}: & =\left\{j\in[1,n]:H(V^j|V^{1:j-1} X^{1:n})>\delta_n \right\} \stepcounter{equation}\tag{\theequation}\label{eq: hv}.
 \end{align*}}
\vspace{-0.2cm}
Partition the set $[1,n]$ into four disjoint sets:
\begin{align*}
& B_1 := \mathcal V_{U|XS} \cap \mathcal H_{U|X}=\mathcal V_{U|XS} , \quad  B_2 : = \mathcal V_{U|XS} \cap \mathcal H_{U|X}^c=\emptyset,\\
& B_3 := \mathcal V_{U|XS}^c \cap \mathcal H_{U|X} ,\quad
 B_4 := \mathcal V_{U|XS}^c \cap \mathcal H_{U|X}^c=\mathcal H_{U|X}^c.
\end{align*}

\begin{oss}\label{oss card2}
We have:
\begin{itemize}
\item[\textbullet] $\mathcal V_{U|XS} \subset \mathcal H_{U|XS}$ and $\displaystyle \lim_{n \rightarrow \infty} \frac{\lvert \mathcal H_{U|XS} \setminus \mathcal V_{U|XS}  \rvert}{n} = 0$ \cite{arikan2010source},
 \item[\textbullet] $B_1  = \mathcal V_{U|XS}$ and $\displaystyle \lim_{n \rightarrow \infty} \frac{\lvert\mathcal V_{U|XS}\rvert}{n}  = H(U|XS)$ \cite {chou2015secretkey},
  \item[\textbullet]  $B_4 = \mathcal H_{U|X}^c$ and $\displaystyle \lim_{n \rightarrow \infty} \frac{\lvert \mathcal V_{U|X}^c \rvert}{n} = 1 - H(U|X)$ \quad \cite {chou2015secretkey},
 \item[\textbullet]  $B_3 \cup\! B_4 \!=\! \mathcal V_{U|XS}^c$ and $\displaystyle \lim_{n \rightarrow \infty} \!\!\frac{\lvert \mathcal V_{U|XS}^c \rvert}{n}\! = \!1 \!- \!H(U|XS)$ \cite {chou2015secretkey}.
\end{itemize}
Note that  $H(U|X)- H(U|XS) = I(X, U;S)\geq 0$ and
 $\lvert B_3 \rvert/n $ tends to $I(X, U;S)$. Since $I(X,U;Y) = I(X;Y)$, the inequality $I(X,U;S) \leq I(X,U;Y)$  implies directly that for $n$ large enough $\lvert B_3 \rvert \leq \lvert A_2\rvert - \lvert A_3 \rvert $.
\end{oss}
\vspace{-0.5cm}

\begin{algorithm}[h!]\label{alg1}
\DontPrintSemicolon
\SetAlgoVlined 
\SetKwInOut{Input}{Input}
\SetKwInOut{Output}{Output}
\Input{ $(S_0^{1:n}, \ldots, S_k^{1:n})$, local randomness (uniform random bits) $M$ and 
common randomness $C=(C_1, K_1, C_2, K_2)$ shared with Node 2:
\begin{itemize}
\item[\textbullet] $C_1$ of size $\lvert A_1 \rvert$ and $K_1$ of size $\lvert A_3 \rvert$;
\item[\textbullet] $C_2$ of size $\lvert B_1 \rvert$ and $K_2$ of size $\lvert B_3 \rvert$.
\end{itemize}}

\Output{$( \widetilde Z^{1:n}_1, \ldots, \widetilde Z^{1:n}_k )$, $( \widetilde V^{1:n}_1, \ldots, \widetilde V^{1:n}_k)$}
\If{$i=1$}{
 $\widetilde Z_{1}[A_1] \longleftarrow C_1 \qquad \widetilde Z_{1}[A_2]  \longleftarrow M $\;
\For{$j \in A_{3} \cup A_{4}$}{Successively draw the bits $\widetilde Z_{i}^j$ according to 
\vspace{-0.2cm}
\begin{small}
\begin{equation} \label{eq: p11}
P_{Z^j \mid Z^{i:j-1}} \left(\widetilde Z_{i}^j\mid \widetilde Z_i^{i:j-1} \right)
\end{equation}
\end{small}
\vspace{-0.5cm}}\;
\vspace{-0.4cm}
 $\widetilde V_{1}[B_1] \longleftarrow C_2$\;
\For{$j \in B_{3} \cup B_{4}$}{
Given $S_1^{1:n}$, successively draw the bits $\widetilde V_{1}^j$ according to 
\vspace{-0.2cm}
\begin{small}
\begin{equation} \label{eq: p21}
P_{V^j \mid V^{1:j-1}X^{1:n}S^{1:n}} \left(\widetilde V_{i}^j\mid \widetilde V_i^{1:j-1} \widetilde X_{i}^n S_{i-1}^n \right)
\end{equation}
\end{small}
\vspace{-0.5cm}}
}

\For{$i=2, \ldots, k$}{
\vspace{-0.4cm}
\begin{align*}
 &\widetilde Z_{i}[A_1] \longleftarrow C_1 \quad   \widetilde Z_{i}[A'_2] \longleftarrow M\\
 &\widetilde Z_{i}[B'_3] \longleftarrow \widetilde V_{i-1}[B_3] \oplus K_2 \quad \widetilde Z_{i}[A'_3] \longleftarrow \widetilde Z_{i-1}[A_{3}] \oplus K_1\\
 \end{align*}\;
 \vspace{-1.2cm}
\For{$j \in A_{3} \cup A_{4}$}{
Successively  draw the bits $\widetilde Z_{i}^j$ according to \eqref{eq: p11}\;
 $\widetilde V_{i}[B_1] \longleftarrow C_2$\;
\For{$j \in B_{3} \cup B_{4}$}{
Successively draw the bits $\widetilde V_{i}^j$ according to \eqref{eq: p21}.
}
}
}
\BlankLine
\caption{Encoding algorithm at Node 1}
\end{algorithm}

\begin{center}\label{fig:kblo}
\begin{figure}[ht!]
 \centering
 \includegraphics[scale=0.7]{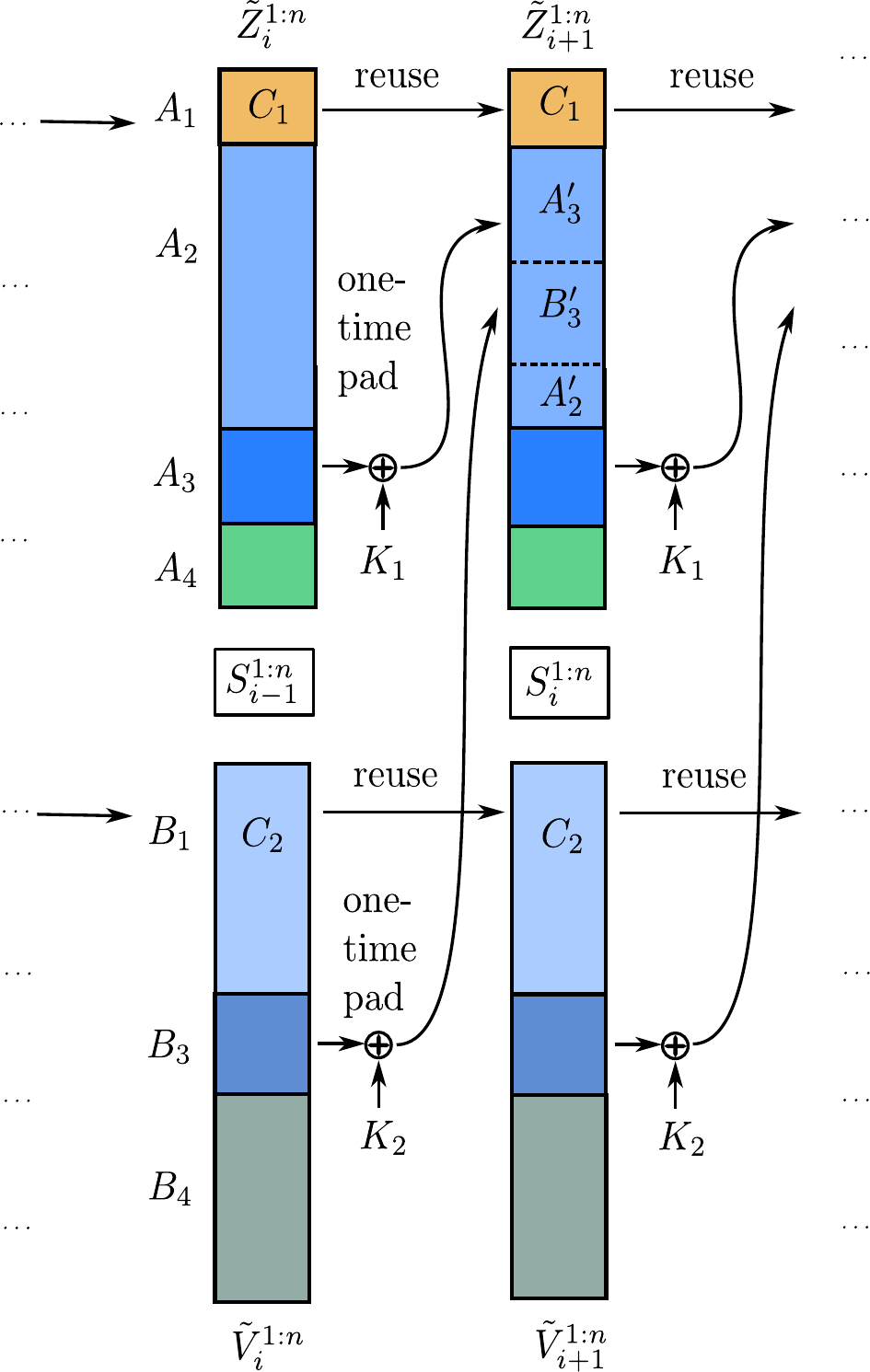}
\caption{Chaining construction for block Markov encoding}
\end{figure}
\end{center}

\begin{algorithm}[h!]\label{alg2}
\DontPrintSemicolon
\SetAlgoVlined 
\SetKwInOut{Input}{Input}
\SetKwInOut{Output}{Output}
\Input{$(Y_1^{1:n}, \ldots, Y_{k+1}^{1:n})$, $C=(C_1,K_1, C_2, K_2)$ common randomness shared with Node 1}
\Output{$(\widehat Z_1^{1:n}, \ldots, \widehat Z_{k}^{1:n})$, $(\widehat V_1^{1:n}, \ldots, \widehat V_{k}^{1:n})$}
\For{$i=k, \ldots, 1$}{
 $\widehat Z_i[A_1] \longleftarrow C_1 \qquad \widehat V_i[B_1] \longleftarrow C_2$\;
 \If{$i=k$}{$\widehat Z_{k}[A_3] \longleftarrow Y_{k+1}^{1:n} \qquad \widehat V_{k}[A_3] \longleftarrow Y_{k+1}^{1:n}$ }
 \Else{
$\widehat Z_{i}[A_3] \leftarrow \widehat Z_{i+1} [A'_{3}] \oplus K_1$ \\
$\widehat V_{i}[B_3] \leftarrow \widehat Z_{i+1} [B'_{3}] \oplus K_2$}
\For{$j \in A_{2} \cup A_{4}$}{ 
Successively draw the bits according to
\vspace{-0.3cm}
\begin{equation*}
 \widehat Z_i^j = \begin{cases}
 0 \quad \mbox{if } L_n(Y_i^{1:n}, Z_i^{1:j-1}) \geq 1\\
 1 \quad \mbox{else} 
 \end{cases}
\end{equation*}
\vspace{-0.5cm}
\begin{small}
 \begin{equation*}
 L_n(Y_i^{1:n}, Z_i^{1:j-1}) = \frac{P_{Z_i^j \mid Z_i^{1:j-1}  Y_i^{1:n}}\left(0 \mid \widehat Z_i^{1:j-1} Y_i^{1:n} \right) }{P_{Z_i^j \mid Z_i^{1:j-1}  Y_i^{1:n}}\left(1 \mid \widehat Z_i^{1:j-1}Y_i^{1:n} \right)}
 \end{equation*}
\end{small}
}
\For{$j \in B_{4}$}{ Successively draw the bits according to
\vspace{-0.3cm}
$$\widehat V_i^j = \begin{cases}
 0 \quad \mbox{if } L_n(X_{i+1}^{1:n}, V_i^{1:j-1}) \geq 1\\
 1 \quad \mbox{else} 
 \end{cases}$$
\vspace{-0.3cm}
}
}
\BlankLine
\caption{Decoding algorithm at Node 2}
\end{algorithm}
\vspace{-0.7cm}
\paragraph{Encoding}
\vspace{-0.5cm}
We use a chaining construction over multiple blocks. 
The encoder observes $(S_0^{1:n}, S_1^{1:n}, \ldots, S_k^{1:n})$, 
where $S_0^{1:n}$ is a uniform random sequence and $S_i^{1:n}$ for $i\in [1, k]$ are $k$ blocks of the source.
It then generates for each block $i\in [1, k]$ random variables $\widetilde Z_i^{1:n}$ and $\widetilde V_i^{1:n}$ following the procedure described in Algorithm 1.
In particular, the chaining construction proceeds as follows.
The bits in  $A_1 \subset \V_{X}$ and $B_1 \subset \V_{U|XS}$ are chosen with uniform probability using uniform randomness sources $(C_1,C_2)$ shared with Node 2, and their value is reused over all blocks.
In the first block the bits in  $A_2 \subset \V_{X}$ are chosen with uniform probability using a local randomness source $M$.
The bits in $A_3 \cup A_4$  and $B_3 \cup B_4$ are generated according to the previous bits using successive cancellation encoding 
\cite{arikan2010source}.
Note that it is possible to sample efficiently from  $P_{Z^j \mid Z^{1:j-1}}$  and
$P_{V^j \mid V^{1:j-1}X^{1:n}S^{1:n}}$ (given $S^{1:n}$ and $X^{1:n}$) respectively \cite{arikan2010source}.

From the second block, let $A'_3$ and $B'_3$ be two disjoint subsets of $A_2$ such that $\lvert A'_3 \rvert= \lvert A_3 \rvert$ and $\lvert B'_3 \rvert= \lvert B_3 \rvert$.  
The existence of those disjoint subsets is guaranteed by Remark \ref{oss card1} and Remark \ref{oss card2}.
The bits of $A_3$ and $B_3$ in block $i$ are used as $A'_3$ and $B'_3$ in block $i+1$ using one time pads with keys $K_1$ and $K_2$ respectively. 
Thanks to the Crypto Lemma  \cite[Lemma 3.1]{bloch2011physical}, if we choose $K_1$ of size $\lvert A_3 \rvert$ and  $K_2$ of size $\lvert B_3 \rvert$ 
to be uniform random keys, the bits in $A'_3$ and $B'_3$ in the block $i+1$ are uniform. The bits in $A'_2:=A_2 \setminus (A'_3 \cup B'_3)$ are chosen with uniform probability using the local randomness source $M$.

The encoder then computes 
$\widetilde X_i^{1:n}=\widetilde Z_i^{1:n} G_n$  for $i=1, \ldots, k$ and sends it over the channel.
We use an extra $(k+1)$-th block to send 
a version of $\widetilde Z_k[A_{3}]$ encoded with a good channel code as in \cite[Section III.B]{Cervia2016}.

\vspace{-0.5cm}
\paragraph{Decoding} The decoder observes $(Y_1^{1:n}, \ldots, Y_{k+1}^{1:n})$ and the $(k+1)$-th block allows it to decode in reverse order.
In block $i \in [1,k]$, the decoder has access to $\widehat Z_i[A_{1} \cup A_{3}]= \widehat Z_i[\mathcal H_{X| Y}]$ and 
$\widehat V_i[B_{1} \cup B_{3}]= \widehat V_i[\mathcal H_{U |X}]$: the bits in $A_1$ and $B_1$ correspond to shared randomness $(C_1, C_2)$,
in block $i \in [1, k-1]$ the bits in $A_3$ and $B_3$ are obtained by successfully recovering $A_2$ in block $i+1$ and
in block $k$ they are recovered from $Y_{k+1}^{1:n}$ as in \cite[Section III.C]{Cervia2016}.
For each block $i=k, \ldots, 1$ the decoder recovers the estimates $\widehat Z_i^{1:n}$ and $\widehat V_i^{1:n}$ using Algorithm 2. 
From ${Y_i}^{1:n}$ and $\widehat Z_{i}[A_1 \cup A_{3}]$ the successive cancellation decoder can retrieve $\widehat Z_i[A_2 \cup A_4]$ and therefore $\widehat V_i[B_4]$.
Note that, as shown in \cite[Theorem 3]{arikan2010source}, $\widetilde{V}^{1:n}$ is equal to  $\widehat V^{1:n} $ with high probability.
The decoder computes $\widehat U_i^{1:n} = \widehat V_i^{1:n} G_n $. 
It then generates $\widehat S_i^{1:n}$ symbol by symbol using: $P_{\widehat S_i^j | \widehat U_i^j Y_i^j} (s|u,y)=P_{\widehat S | UY}(s|u,y).$
 
\begin{oss}
The rate of common randomness is negligible, since:
\begin{align*}
 &\lim_{n \rightarrow \infty \atop k \rightarrow \infty} \frac{  \lvert A_1 \rvert +   \lvert A_3 \rvert + \lvert B_1  \rvert +  \lvert B_3 \rvert }{kn}  = \lim_{n \rightarrow \infty \atop k \rightarrow \infty}\frac{ \lvert \mathcal V_{X|Y} \rvert + \lvert \mathcal H_{U|X}  \rvert}{nk}\\
 &= \lim_{k \to \infty} \frac{H(X|Y)+ H(U|X)}{k} =0.
\end{align*}
\end{oss}

\vspace{-0.5cm}
\section{Proof of Theorem \ref{theo}}\label{subsec:coord}
Given $\varepsilon >0$, we want to prove that: 
$$\lim_{n \rightarrow \infty} \mathbb P \left\{ \mathbb V\left(T_{S^{1:n}_{1:k+1} X^{1:n}_{1:k+1} Y^{1:n}_{1:k+1} \widehat{S}^{1:n}_{1:k+1}}, P_{SXY\widehat{S}}\right) > \varepsilon \right\} =0.$$
This requires a few steps:\\
\textbf{1.} $\forall  i\!\in\!  \![1,k ]$,
$\displaystyle \!\lim_{n \rightarrow \infty} \!\mathbb P \{ \mathbb V(T_{S^{1:n}_{i}\! \widetilde X^{1:n}_{i} \widetilde U^{1:n}_{i}\! }, P_{SXU})\! > \!\varepsilon \}\! =0;$\\
\textbf{2.} $\forall  i\!\in\!  \![1,k ]$,
$\displaystyle \!\lim_{n \rightarrow \infty} \!\mathbb P \{ \mathbb V(T_{S^{1:n}_{i}\! \widetilde X^{1:n}_{i} \widetilde U^{1:n}_{i} Y^{1:n}_{i}}, P_{SXUY})\! > \!\varepsilon \}\! =0;$\\
\textbf{3.} $\forall  \!i\!\in\!  \![1,k \!]$,
$\displaystyle \!\!\lim_{n \rightarrow \infty} \!\mathbb P \{ \mathbb V(T_{S^{1:n}_{i}\! \widetilde X^{1:n}_{i} \widetilde U^{1:n}_{i} Y^{1:n}_{i} \!\hat{S}^{1:n}_{i}},P_{\!SXUY\hat{S}})\!\! > \!\!\varepsilon \}\!\! =\!0;$\\
\textbf{4.} Convergence in each block implies overall convergence;\\
\textbf{5.} The theorem follows from the fact that 
\begin{align*}
&\mathbb V \left(T_{S^{1:n}_{1:k+1} \widetilde X^{1:n}_{1:k+1}  Y^{1:n}_{1:k+1} \widehat{S}^{1:n}_{1:k+1}},P_{SXY \widehat S} \right) \leq \\
&\mathbb V(T_{S^{1:n}_{1:k+1} \widetilde X^{1:n}_{1:k+1} \widetilde U^{1:n}_{1:k+1} Y^{1:n}_{1:k+1} \widehat{S}^{1:n}_{1:k+1}},P_{SXUY \widehat S}).
\end{align*}

Note that since the steps 2 to 5 have already been proved in \cite[Section IV]{Cervia2016}, we only need to prove the first step.
For all $\varepsilon_0 >0$, we define
{\allowdisplaybreaks
\begin{align*}
  & \mathcal{T}_{\varepsilon_0}\left( P_{SXU} \right) := \left\{ (\mathbf s,\mathbf x, \mathbf u) \big| \mathbb{V}\left( P_{SXU}, T_{(\mathbf s,\mathbf x, \mathbf u)} \right) \leq \varepsilon_0\right\}
  \end{align*}}
Observe that for the i.i.d. distribution, we have $ \lim_{n \rightarrow \infty} \mathbb P \left\{  (\mathbf s,\mathbf x, \mathbf u) \in \mathcal{T}_{\varepsilon_0}\left( P_{SXU} \right) \right\}=1.$

Let $i \in [1,k]$, we have:
{\allowdisplaybreaks
\begin{align*}
& \mathbb P \left\{ \mathbb{V}\left( T_{S^{1:n}_i X^{1:n}_i U^{1:n}_i} , P_{SXU}\right) > \varepsilon_0\right\}\\
& = \sum_{\mathbf s,\mathbf x, \mathbf u} P_{S^{1:n}_i \widetilde X^{1:n}_i \widetilde U^{1:n}_i} \left(\mathbf s,\mathbf x, \mathbf u \right)\mathds{1}\left\{ (\mathbf s,\mathbf x, \mathbf u) \notin  \mathcal{T}_{\varepsilon_0}\left( P_{SXU} \right) \right\}\\
& =  \sum_{\mathbf s,\mathbf x, \mathbf u} ( P_{S^{1:n}_i  \widetilde X^{1:n}_i \widetilde U^{1:n}_i} \left(\mathbf s,\mathbf x, \mathbf u \right) - P _{S^{1:n}X^{1:n} U^{1:n}} \left(\mathbf s,\mathbf x, \mathbf u \right) \\
& +  P _{S^{1:n}X^{1:n} U^{1:n}} \left(\mathbf s,\mathbf x, \mathbf u\right) ) \mathds{1}{\left\{ (\mathbf s,\mathbf x, \mathbf u) \notin  \mathcal{T}_{\varepsilon_0}\left( P_{SXU} \right) \right\}}\\
& \leq \! \mathbb{V} ( P_{S^{1:n}\! \widetilde X^{1:n} \!\widetilde U^{1:n}}, P_{S^{1:n}\!X^{1:n}\! U^{1:n}}) \!+ \!\mathbb P\{ \! (\mathbf s,\!\mathbf x,\! \mathbf u\! ) \notin \mathcal{T}_{\varepsilon_0}( P_{S\!X\!U}) \}
\end{align*}}
which tends to 0 thanks to a typicality argument and the following result.

\begin{lem}\label{dis}
For $i \in [1,k]$, let $\delta_n = 2^{-n^ {\beta}}$ where $0<\beta<1/2$,
 $$\mathbb{V} \left({P}_{S_i^{1:n} \widetilde X_i^{1:n} \widetilde U_i^{1:n}}, P_{S^{1:n} X^{1:n} U^{1:n} } \right) \leq 2 \sqrt{ \log2} \sqrt{n \delta_n}.$$
\begin{proof}[Proof]
By the chain rule, we have
{\allowdisplaybreaks
\begin{align*}
& \mathbb{D}\left( P_{S^{1:n} X^{1:n} U^{1:n}} \Big\Arrowvert  P_{S_i^{1:n} \widetilde X_i^{1:n} \widetilde U_i^{1:n}}\right) \stepcounter{equation}\tag{\theequation}\label{eq: div}\\
& = \mathbb{D}\left(  P_{X^{1:n}| S^{1:n}} \Big\Arrowvert P_{\widetilde X_i^{1:n}| S_i^{1:n}}  \Big| P_{S^{1:n}} \right)\\
&+  \mathbb{D}\left(P_{U^{1:n} |X^{1:n} S^{1:n}} \Big\Arrowvert   P_{\widetilde U_i^{1:n}| \widetilde X_i^{1:n} S_i^{1:n}}\Big| P_{X^{1:n} S^{1:n}}  \right)
\end{align*}}
We call $D_1$ and $D_2$ the first and the second term.
Then: 
{\allowdisplaybreaks
\begin{align*}
D_1 & {\overset{{(a)}}{=}} \mathbb{D}\left(  P_{X^{1:n}} \Big\Arrowvert P_{\widetilde X_i^{1:n}} \right) {\overset{{(b)}}{=}} \mathbb{D}\left(  P_{Z^{1:n}} \Big\Arrowvert P_{\widetilde Z_i^{1:n}} \right) \stepcounter{equation}\tag{\theequation}\label{divA}\\
& {\overset{{(c)}}{=}} \sum_{j=1}^n \mathbb{D}\left( P_{ Z_i^{j}|  Z_i^{1:j-1}}   \Big\Arrowvert   P_{\widetilde Z_i^{j}| \widetilde Z_i^{1:j-1}}\Big| P_{ Z_i^{1:j-1}} \right)\\
& {\overset{{(d)}}{=}} \sum_{j \in A_1 \cup A_2} \mathbb{D}\left(  P_{ Z_i^{j}|  Z_i^{1:j-1}}   \Big\Arrowvert P_{\widetilde Z_i^{j}| \widetilde Z_i^{1:j-1}} \Big| P_{ Z_i^{1:j-1}}\right)\\
& {\overset{{(e)}}{=}} \sum_{j \in A_1 \cup A_2} \left(1- H\left(Z_i^{j}|  Z_i^{1:j-1}\right)\right)  {\overset{{(f)}}{<}} n \lvert \mathcal V_X \rvert \leq n \delta_n
\end{align*}}
where $(a)$ comes from the fact that $X$ is independent of $S$, $(b)$ from the invertibility of $G_n$,  $(c)$ from the chain rule, $(d)$ from \eqref{eq: p11}, $(e)$ from the fact that the conditional distribution $P_{\widetilde Z_i^{j}| \widetilde Z_i^{1:j-1}}$ is uniform for $j$ in $A_1$ and $A_2$ and $(f)$ from Definition \eqref{eq: hz}.

Similarly, $D_2 <n \delta_n $. Then $D_1+D_2 < 2 n \delta_n$ and the proof is completed using Pinsker's inequality. 

\end{proof}
\end{lem}

\begingroup
    \setlength{\bibsep}{5pt}
    \setstretch{0.9}
\begin{footnotesize}
 \bibliographystyle{IEEEtran}
 \renewcommand\refname{References}
\bibliography{mybib}
\end{footnotesize}
\endgroup

\end{document}